\documentclass[conference,letter,comsoc]{IEEEtran}

\usepackage{xspace}
\usepackage{balance}
\usepackage{caption,subcaption}
\usepackage[skins]{tcolorbox}
\usepackage[inline]{enumitem}
\usepackage{amsmath}
\usepackage{graphicx}
\usepackage{makecell}
\usepackage{colortbl}
\usepackage{hyperref}
\usepackage{xspace}

\definecolor{darkgreen}{rgb}{0.0, 0.3, 0.13}
\definecolor{darkred}{rgb}{0.2, 0.0, 0.13}
\definecolor{darkred2}{rgb}{0.7, 0.04, 0.04}
\definecolor{darkYellow}{rgb}{0.9, 0.25, 0.04}
\definecolor{darkblue}{rgb}{0.04, 0.44, 0.7}
\newcommand{\cc}[1]{\cellcolor{darkred2!50!white!#1}}
\newcommand{\dd}[1]{\cellcolor{darkblue!50!white!#1}}

\newcommand{\oo}[1]{\cellcolor{darkYellow!50!white!#1}}

\newcommand{\etal}{{\em et al}.\xspace}

\newcommand{\BfPara}[1]{\vspace{1mm}{\noindent\bf #1.}\xspace}

\begin{document}

\title{Understanding the Country-Level Security of Free Content Websites and their Hosting Infrastructure}

\author{
\IEEEauthorblockN{Mohammed Alqadhi$^{\diamond,1}$, Ali Alkinoon$^{\diamond,2}$, Saeed Salem$^{\dagger,3}$ and David Mohaisen$^{\diamond,4}$}
\IEEEauthorblockA{$^\diamond$University of Central Florida \hspace{10mm} $^\dagger$Qatar University\\
\{{\tt $^1$malqadhi, $^2$aalkinoon, $^4$mohaisen}\}@{\tt ucf.edu} \hspace{10mm} {\tt $^3$saeed.salem}@{\tt qu.edu.qa}}
}

\maketitle
\begin{abstract}
This paper examines free content websites (FCWs) and premium content websites (PCWs) in different countries, comparing them to general websites. The focus is on the distribution of malicious websites and their correlation with the national cyber security index (NCSI), which measures a country's cyber security maturity and its ability to deter the hosting of such malicious websites. By analyzing a dataset comprising 1,562 FCWs and PCWs, along with Alexa's top million websites dataset sample, we discovered that a majority of the investigated websites are hosted in the United States. Interestingly, the United States has a relatively low NCSI, mainly due to a lower score in privacy policy development. Similar patterns were observed for other countries With varying NCSI criteria. Furthermore, we present the distribution of various categories of FCWs and PCWs across countries. We identify the top hosting countries for each category and provide the percentage of discovered malicious websites in those countries. Ultimately, the goal of this study is to identify regional vulnerabilities in hosting FCWs and guide policy improvements at the country level to mitigate potential cyber threats.
\end{abstract}


%\maketitle

\section{Introduction} \label{sec:Introduction}
Free content websites (FCWs) provide free content to their users, including books, games, music, movies, and software. The same type of content can be provided to the user at a cost in premium content websites (PCWs). In the prior work~\cite{AlabduljabbarM22}, the privacy policies of FCWs were shown to be less elaborate compared to PCWs, where FCWs reuse policies depending on their hosting providers. These hosting providers usually reside in one or more countries where users can access the FCWs and PCWs. 

It is important to assess the country-level (geo-distribution) security features of FCWs and PCWs to understand their ecosystem and provide the appropriate security recommendations. Moreover, as the characteristics of those websites may differ based on the type of content they provide, a per-category analysis is paramount for deeper contrasts. Finally, as the target of analysis at the country level, it is essential to understand the security of such websites and country security matureness---e.g., measured by the national cyber security index (NCSI). 

\BfPara{Why Study the Geographical Distribution} By investigating the geo-distribution of FCWs, we can design better strategies to ensure network security based on the country's regulations and cyber security policies. Understanding the distribution of FCWs over countries is necessary to identify the concentration of malicious websites and infrastructures. As a result, it will help users of these services protect themselves from being victimized by a vulnerability that cannot be controlled or governed by the law of the same country. For example, if a network user uses a FCW that resides in a different country and the website victimizes the user, the user would know if the law of the FCW country is strict and elaborated so that the user can perform legal action against that FCW. They can legally remove any acquisitions, definitions, misinformation, or malicious content from the FCW. Also, it will help to determine the necessary actions against the hosting providers that contain the most malicious websites. 

\BfPara{Why Study the Cyber Security Policy} To understand the factors that improve security at the country level, we investigate the maturity of cyber security policies of countries where most FCWs are hosted. Security agreements could exist between countries with the least malicious websites, which indicates the effectiveness of such policies and agreements.  Investigating the matureness of the cyber security policies of the countries would reveal if the countries with the most concentration of malicious FCWs are those with the least mature cyber security policies. Knowing the NCSI for each country will help us find if there is any correlation between the actual security landscape measured by the prevalence of those websites and the cyber security policies.

Our results show certain country-level distribution patterns for FCWs and PCWs, where most of the malicious websites are heavily concentrated in some countries. Similar conclusions are derived from the results of the general websites with some differences in concentration where the NCSI average shows a unique pattern for the top hosting countries, where we notice lower scores in some digital development aspects in the countries that contribute to hosting a higher rate of the malicious FCWs, where we notice the significant difference in the distribution of the category websites analysis results between every category and between FCWs and PCWs in general.

\BfPara{Contributions} Using 1,562 FCWs and PCWs~\cite{AlabduljabbarM22, AlabduljabbarAMM21, AlqadhiATSNM22}, we contribute the following. 
\begin{enumerate*}
    \item {\em Malicious FCWs Measurements.} We identify the malicious FCWs and PCWs and analyze their connections to various infrastructure entities and characteristics. More specifically, we revealed the countries contributing most to (malicious) FCW hosting.

     \item {\em Comparative Analysis.} A thorough comparison was made between FCWs and PCWs in terms of their utilization of infrastructure and security features in every hosting country. Furthermore, the top hosting countries for Alexa's one million websites were compared to the  FCWs and PCWs and maliciousness within each hosting entity. This yielded a precise, inclusive, and contextualized description of FCWs when put next to PCWs.

     \item {\em Per-category Analysis.} We performed a comprehensive analysis of the most contributed hosting countries for the different content FCWs and PCWs categories. We give a detailed comparison in every content type between FCWs and PCWs, describing the affinities for every studied category. 

     \item {\em Security Policy Impact.} We investigated the correlation analysis that sheds light on the average NCSI score of the top FCWs and PCWs hosting countries to identify the role of the privacy policy development on the percentage of the malicious hosted FCWs in that country.
\end{enumerate*}

\BfPara{Paper Organization} In section~\ref{sec:related}, we present the related work. In section~\ref{sec:Methodology}, we describe the data collection and analysis dimensions. In section~\ref{sec:results}, we present our findings. In section~\ref{sec:discussion}, we provide our discussion. In section~\ref{sec:final}, we present our concluding remarks.

\section{Related Work}\label{sec:related}

Several works analyzed the security characteristics of FCWs such as~\cite{AlabduljabbarAMM21, AlabduljabbarMAJCM22, AlabduljabbarMCJCM22, AlabduljabbarM22,AlqadhiATSNM22, HuT17, HuTB19, LeeNHC22, RoyKN22}, while other works focus on analyzing the security of the general websites~\cite{CalzavaraRB16, CalzavaraRB18, DobolyiA16, EnglehardtN16, GhaffarianS17, KontaxisAPM11, KosbaMWTK14, LiZXYW12, Libert15, MaticTS19, Mohaisen15, RaponiP20, SamarasingheAMY22, ShimamotoYOCOO19, WashRBW16, WickramasingheNTKZ21}. Moreover, several works explored the regional analysis for domain-specific websites, such as governments and universities. The key difference in this study is that we investigate the security of FCWs across different countries, utilizing various new features of their modeling and contrasting them to the general web population of websites. This, as a result, supplements the other efforts focused on understanding the security, privacy policies, accessibility, or performance of such websites~\cite{AlabduljabbarAMM21, AlabduljabbarM22, BangeraG17, Sandoval-Guzman17, shafqatM16, VaughanZ07, VelasquezE18, VerkijikaW18, WakelingKJKS22, ZareR14, ZhaoZZ10, MacakogluM22}.
Given the multitude of studies and space constraints, we focus on a select group of highly relevant studies concerning this work and findings.

%\BfPara{Free Content Websites Analysis} 
In a previous study, Alabduljabbar \etal~\cite{AlabduljabbarM22} investigated FCWs' privacy policies and their expressiveness utilizing TLDR~\cite{AlabduljabbarAMM21}, a natural language processing (NLP) pipeline for privacy policy analysis. They also examined the uniqueness of the policy for each FCW compared to PCWs. Among other interesting findings, they concluded that FCWs' privacy policies are unclear in stating their data collection practices do not provide useful information compared to PCWs' policies, and utilize mostly predefined privacy policy templates, which may not state the actual data tracking, storage, and sharing practices of users data.

%\BfPara{General Websites Analysis}
Another study investigated the security of the top-used website lists provided by Alexa. Raponi and Di Pietro~\cite{RaponiP20} performed a detailed analysis of Alexa's top 200 websites with domains registered in certain European countries and analyzed the password recovery management mechanisms adopted by each website. They found more than 54\% of the websites in France, 36\% in Italy, 47\% in Spain, and 33\%  in the United Kingdom to be vulnerable in  December 2017, with almost no difference a year later, highlighting minimal progress in adapting the General Data Protection Regulation (GDPR) standard. Verkijika \etal~\cite{VerkijikaW18} investigated the public values delivery of web-based platforms in Sub-Saharan Africa by analyzing 279 e‐government websites from 31 countries, revealing a lack of various features associated with accessibility, citizen engagement, trust development, responsiveness, dialogue, and quality of service among the surveyed websites.

Shafqat \etal~\cite{shafqatM16} conducted a comparative analysis of 20 countries' National Cyber Security Strategies (NCSS) using different metrics, such as perception of cyber threats, organization overviews, critical sectors and infrastructure, incident response capabilities, etc. Their results show that while all countries have defensive measures to protect their cyberspace from threats, there is variation in the approaches they use. The study concludes with recommendations nations may use to design their NCSS documents with global best practices for improved cyber resilience.

Vaughan~\etal~\cite{VaughanZ07} examined the coverage of websites from four countries---United States, China, Singapore, and Taiwan---in four major search engines---Google, Yahoo!, MSN, and Yahoo! China---and found that the United States-based sites had higher coverage rates than those from other countries. Moreover, they found that the Chinese sites had the lowest average coverage rate. They also found that the
language factor did not explain this difference in representation, although the visibility, as measured by the number of links to a site, did affect its chance of being indexed. Yahoo! China provided better coverage for Chinese and surrounding region sites than the global Yahoo! search engine. 

Velasquez~\etal~\cite{VelasquezE18} investigated the accessibility, resources, and staff availability provided by 1,517 public library websites in Australia, Canada, and the United States. The research aimed to extend the definition of physical library branches into their digital counterparts. To assess this, 18 criteria were used to determine if they were present on each website, and descriptive statistics revealed that Canadian and United States libraries met more criteria than Australian libraries. However, many similarities between all three countries' websites were found overall. Bangera~\etal~\cite{BangeraG17}  presents the results of an extensive study of web hosting, with a particular focus on differences between ads and regular content. A virtual private network (VPN) service was used to collect data from top country-specific websites in 52 countries, and the findings show that ads employ more servers for broader load distribution. At the same time, replication is local for ads and global for regular content. 

While there is an overlap between the prior work and ours in the analyzed modalities, our work stands out in utilizing those modalities to understand the ecosystem of free content websites with the premium and general websites to their geographical distribution and co-location in countries with varying security policy standings.  
\section{Methodology}\label{sec:Methodology}

\subsection{Research Questions}
This work aims to understand how FCWs are hosted for their geographic locations and the divergence from PCWs and general website populations. To accomplish our goal, we endeavor to address several questions. {\bf RQ1}. Where are malicious websites mainly concentrated, and what correlations exist regarding their geographical locations? {\bf RQ2}. Are there any similarities or differences between FCWs, PCWs, and the general websites in their geographical distribution regarding the malicious websites? {\bf RQ3}. What are the main geographical distribution characteristics of the different content category websites? {\bf RQ4}. Are there any inconsistencies between top countries hosting malicious websites and the national cyber security index (NCSI) of those countries?

\subsection{Dataset and Data Collection}
We use several datasets: \begin{enumerate*}
\item A primary dataset consisting of FCWs, PCWs, and their corresponding annotations.
\item A dataset for the general website population to facilitate our contrast evaluation between FCWs, PCWs, and their utilization of infrastructure. 
\item The results of network features extraction using ipdata~\cite{ipdata} and IPSHU~\cite{ipshu} in order to obtain the country-level annotation for every website in the scanned dataset.
\item The results of malicious annotation websites by scanning all websites datasets using VirusTotal~\cite{VirusTotal}. In the following, we will review those datasets and how we obtained them.
\end{enumerate*}

\subsubsection{Free and Premium Content Websites Dataset}
Our study employs a dataset of 1,562 websites compiled as per the criteria established in the prior works~\cite{AlabduljabbarM22, AlabduljabbarMAJCM22, AlabduljabbarMCJCM22, AlqadhiATSNM22}. When determining whether to add a website to this list, the primary considerations are its level of popularity, the language used on the site, and how active it is. To assess {\em popularity}, upon entering a keyword, the ranking of each website on major search engines is assessed. Websites that utilize English as their primary language are kept for additional examination. Meanwhile, {\em activity} is determined by verifying that the website returned from the search engine is online (active) at the time of the evaluation. 

Similarly, three search engines---Google, DuckDuckGo, and Bing---are employed to estimate website popularity, where the average rank of the returned website is taken in making this estimation. Manual inspection is employed in determining whether a website is an FCW or PCW. Moreover, each webpage is assigned a category based on the content it features---books, games, movies, music, and software. Finally,  the classification of websites and other relevant keywords (e.g., free, premium, paid, etc.) are used for searching in the related search engine. Upon obtaining the filtered websites from the previous steps, we initiated a query to their domain names to acquire their hosts' IP addresses. We found that 1,509 websites were available, which represents 96.6\% of the total websites we queried. Among these active websites, 788 were FCWs, and 721 were  PCWs. Based on the content type, we categorize the free and premium websites into five categories: books (144 free and 191 premium), games (78 free and 111 premium), movies (310 free and 152 premium), music (80 free and 86 premium) and software (176 free and 181 premium).

\subsubsection{General Websites Sample} 
While measuring and characterizing the free and premium content websites in isolation may shed light on their characteristics, contrasting them with general websites sample can put them into perspective. To this end, we collect a benchmark dataset that is representative of the general websites population. To obtain an unbiased random sample, 2,400 websites were drawn from Alexa's Top One Million website dataset~\cite{AlexaWebsites}. The sample size is selected to be comparable to the total number of websites in the free and premium content websites while ensuring a small error and high confidence when considering the mean estimation of the overall population of the websites. In particular, we used a margin of error of 2\% and a confidence interval of 95\%, resulting in a sample size of 2,400. We note that the size of the population (1 million in this case) has an insignificant effect on the size. We refer to this dataset as the ``general'' for simplicity in presenting the subsequent results.

As in the preprocessing and augmentation of FCWs and PCWs, we consider whether a general website is active or not---i.e., online or offline at the time of the data acquisition. We found that only 2,057 websites were active, which accounts for only 85.7\% compared to 96.5\% for the final dataset of FCWs and PCWs. We obtained each sample's IP address and hosting countries with the {\em ipdata} API in the free and premium content websites. 

\subsubsection{Malicious Websites Annotation}\label{sec:data:maliciousness}
The main goal of this research is to assess the regional concentration of malicious FCWs in comparison to the PCWs and the general websites. To begin, we took advantage of VirusTotal~\cite{VirusTotal}, a service that combines more than 70 scanning engines and can be used to classify whether a domain name (URL), IP address, or binary (file)---identified by its unique hash value---is malicious or benign. Upon passing a file to VirusTotal, it returns a list of antivirus scanners and their associated detections. Our datasets of the different types of websites are further enhanced using the annotation provided by VirusTotal, where we consider a website to be malicious if at least one of the returned results by VirusTotal is malicious and benign otherwise. 

\subsubsection{The National Cyber Security Index} 

The national cyber security index (NCSI)~\cite{ncsi} is provided by the \href{https://ega.ee/}{e-governance academy} and identifies a rating of countries based on 12 metrics: 
\begin{enumerate*}
    \item cyber security policy of that country, 
    \item identified and analyzed security threats, 
    \item education and professional development, 
    \item contribution to global cyber security, 
    \item protection of digital services, 
    \item protection of essential services, 
    \item electronic identification and digital signature, 
    \item protection of personal data, 
    \item cyber incident response, 
    \item cyber crisis management, 
    \item cyber crimes fighting, and 
    \item military cyber operation
\end{enumerate*}

We use NCSI to understand the role of cyber security policy and its association with favorable security outcomes, such as the lack of malicious websites in a given country. We hypothesize that countries with a high NCSI would have a low percentage of such malicious websites, and we examine this hypothesis through correlation analysis. Our justification is that countries with the most malicious FCWs have less mature cyber security policies or may not be aware of the latest cyber threats. This is consistent with the rationale of developing NCSI based on objective qualities of the cyber security posture at the country level, as those countries with a lower rating in NCSI might not be digitally well developed to analyze the recent threats and implement the analysis outcomes into defenses for taking down such websites, or may not protect digital services by applying a high-quality standard and providing a competent supervisory authority that tracks private or public digital services on both FCWs and PCWs. 

To sum up, low-ranked countries may: 
\begin{enumerate*}
    \item have less mature cyber security policies,
    \item lack of awareness of the cyber security threats,
    \item not using recent technology to identify and analyze the risk,
    \item provide less protection of digital services, and
    \item provide low protection of personal data.
\end{enumerate*}

\subsection{Analysis Dimensions}
\begin{figure}[t]
    \centering
    \includegraphics[width=0.98\linewidth]{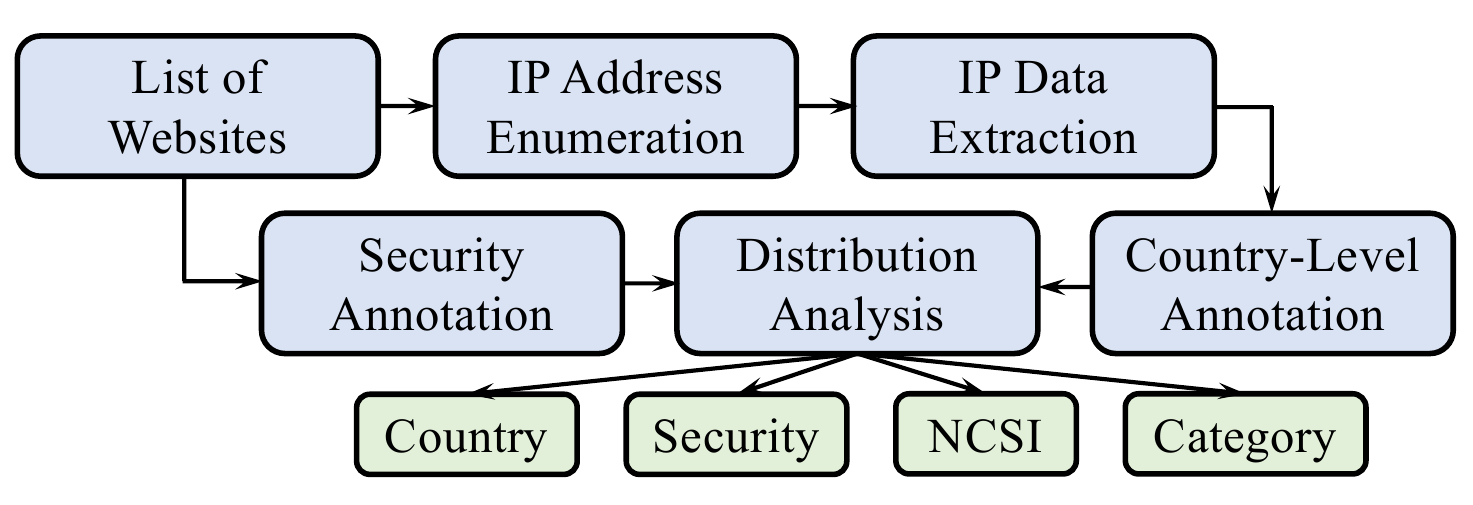}
    \caption{Feature extraction and data enumeration workflow, along with comprehensive distribution analysis leading to the FCWs country-level analysis results.}
    \label{fig:Analysis_Workflow}
\end{figure}

We conduct analyses to detect patterns and disparities between FCWs, PCWs, and general websites across six categories: the country of origin, the number of websites (count), the proportion of the studied websites per each country (percentage), the total maliciousness contribution (malicious count and malicious percentage), and the maliciousness of the hosted websites per feature count and percentage. Each studied dimension is defined below, and the workflow of our analysis pipeline is shown in Figure~\ref{fig:Analysis_Workflow}.

\BfPara{Country} The country where the IP-based infrastructure of FCWs, PCWs, and general websites is located. Our investigation discovered that this attribute has 41 different values that correspond to as many countries.

\BfPara{Count} The total number of websites hosted on an IP address that is located in that particular country and associated with the given class of websites.

\BfPara{Percentage} The number of websites (FCWs, PCWs, general) in a  country normalized by the total number of investigated websites for that type of the studied class of websites. This feature is employed to comprehend any differences in website distribution across our sample.

\BfPara{Malicious Count (MC)} A count of the discovered malicious websites by VirusTotal as described in section~\ref{sec:data:maliciousness} that are hosted within a specific country.

\BfPara{Malicious Per Country Percentage (MPCP)} The proportion of malicious websites relative to the number of websites in a given country. This feature emphasizes the contribution of the studied entity to the total number of malicious websites by considering their size within our dataset. As opposed to MC, which gauges an entity's overall contribution to the total malicious hosting contribution according to our analysis, MPCP normalizes this value by considering how many potentially malicious websites reside within a particular country, acknowledging that countries may vary significantly in their size (scale). 

\BfPara{Malicious Percentage (MP)} This feature indicates the ratio of MC to all websites in the country under analysis for that particular sample, meaning the total number of malicious websites in FCWs, PCWs, both, or general websites. Unlike MC’s indication, MPCP implies that even large entities may contribute little to the total number of malicious websites when their size is considered.

\BfPara{The National Cyber Security Index}\label{sec:ncsi} Provided by NCSI, we studied the following features that determine the high malicious percentage and the weaknesses in NCSI scores:
\begin{enumerate*}
\item {\bf country}, which is the name of the country reported being one of the most hosting FCW/PCW countries, 
\item {\bf count}, the number of websites found in the given country,
\item {\bf MPCP}, the percentage of the malicious websites, as described earlier, discovered per each country,
\item {\bf MP}, normalizes the number of malicious websites in every country over the total number of websites,
\item {\bf NCSI}, which is the National Cyber Security Index, defined earlier,
\item {\bf DDL}, index signifies the Digital Development Level, 
\item {\bf CSPD}, index signifies the Cyber Security Policy Development, 
\item {\bf CTAI}, index signifies the Cyber Threat Analysis and Information,
\item {\bf PDS}, index signifies the Protection of Digital Services,
\item {\bf PPD}, index signifies the Protection of Personal Data, and
\item {\bf Average}, which is the average score for each of the previous features.
\end{enumerate*}

\section{Analysis Results} \label{sec:results}
We provide the results of our findings through an analysis pipeline used to examine the distribution of FCWs and PCWs. We describe and compare trends in FCWs and PCWs distribution across countries, and their comparison to the concentration of the general websites. We will compare the results from each feature analysis we performed, followed by an analysis that takes into account such characteristics with regard to the type of content being hosted on the given website -- i.e., content category analysis (books, games, movies, music, and software). Finally, we provide the NCSI scores analysis, which may reveal some hosting affinities or correlations between the hosting countries and the hosted FCWs and PCWs.

\subsection{Country-Level Distribution Analysis}\label{sec:cc-distribution}

\BfPara{General Insights} The results of the distribution analysis of FCWs and PCWs over different countries reflect the following insights followed by the analysis tables results:
\begin{enumerate*}
\item More than half of the websites reside in the United States, and 33.6\% of the FCWs and PCWs are malicious, as it appears in Table~\ref{tab:FPCountries}. This may imply that putting security measurements on the websites in the US would improve website security by almost 20\% for both FCWs and PCWs.
\item Preventing FCWs from being deployed in Belgium will contribute to changing the classification of Belgium from the second top in hosting malicious websites to exclusively hosting benign websites.
\item Compared to locations globally, FCWs and PCWs in all categories are primarily hosted in the US rather than the rest of the countries.
\end{enumerate*}

\begin{figure}[t]
    \centering
    \includegraphics[width=1\linewidth]{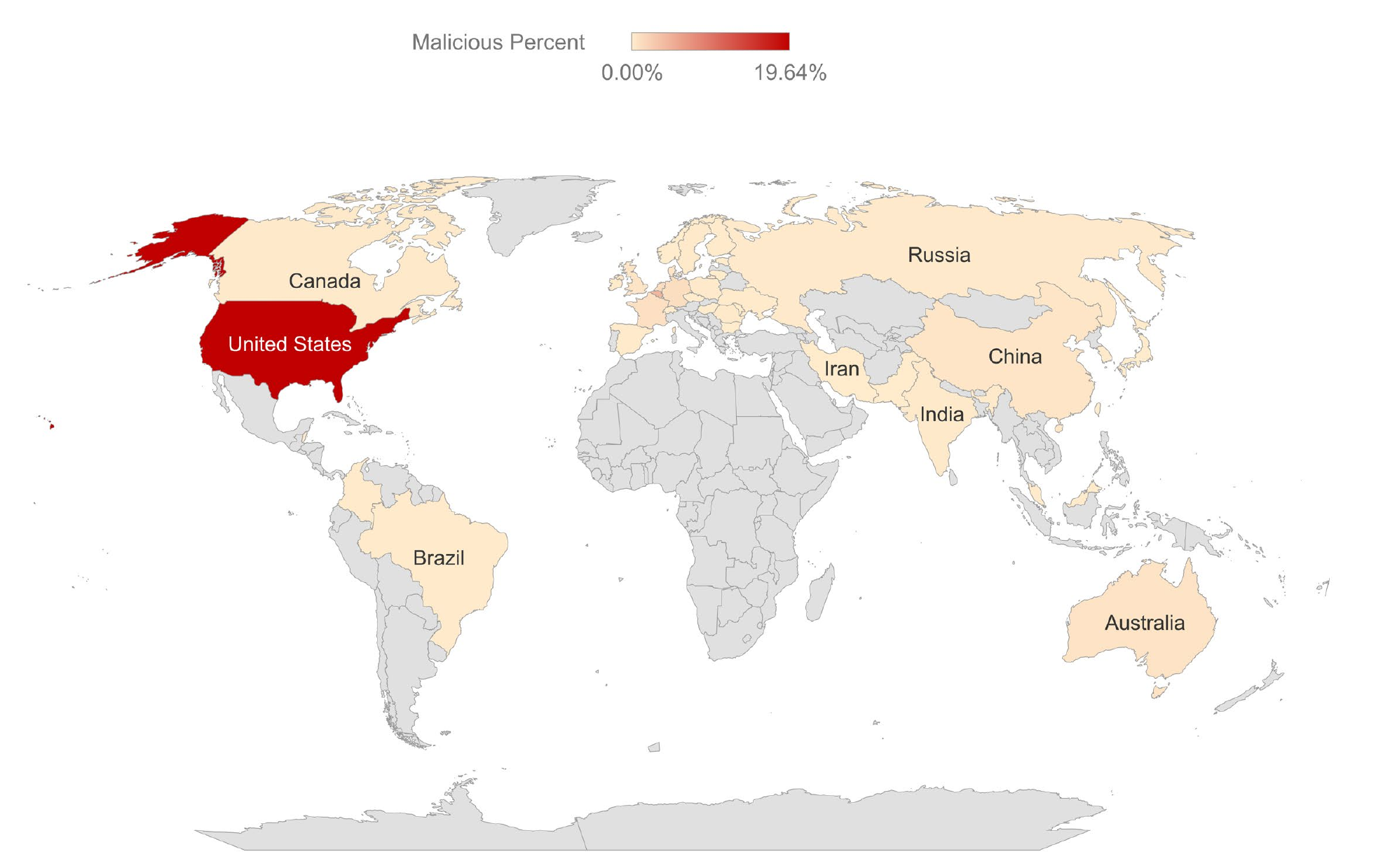}
    \caption{This heatmap illustrates the distribution of malicious websites across various countries.}
    \label{fig:Malicious_Websites}
\end{figure}

\BfPara{Free and Premium Websites} 
Table~\ref{tab:FPCountries} shows the distribution of FCWs and PCWs, along with their respective MC, MPCP, and MP. The United States leads with 58.5\% of the total websites, followed by Belgium at 6.6\%, the Netherlands at 6.3\%, and Germany at 5.9\%. Australia, the United Kingdom, France, China, Canada, and Ireland contribute to the overall distribution, as shown in Figure~\ref{fig:Malicious_Websites}.

In terms of MC, the United States has the highest count of malicious websites (297), followed by Belgium (67) and the Netherlands (19). Moreover, the United States, Belgium, and France exhibit significantly higher MPCP for FCWs, while the other countries maintain relatively lower percentages across all categories. Overall, out of 1,509 websites, 479 were malicious. These results shed light on the prevalence of malicious content across different countries and website categories, offering valuable insights into potential areas of focus for cyber security efforts.

\BfPara{Benchmark Websites} Table ~\ref{tab:MillionCountries} shows the distribution of the general websites across countries for FCWs and PCWs. The United States leads, hosting 45.8\% of the websites, followed by Germany (7.1\%), France (4.9\%), China (3\%), the Netherlands (3\%), and Canada (2.5\%). The United Kingdom, Australia, Ireland, and Belgium have a smaller presence. In terms of MC, the United States has the highest count (44). The overall MP for the Top One Million websites is 4.5\%. The ``Others'' category accounts for 29.5\% of the websites and has an MC of 121. 

Table~\ref{tab:FPCountries} shows the United States also leads in hosting FCWs and PCWs, with $\approx$59\%. Belgium, the Netherlands, and Germany follow, but with much smaller percentages. The distribution of malicious content is more pronounced in the United States, with a considerably higher MP of 19.6\% compared to the Top One Million websites. The MPCP values differ across countries.

In summary, the United States is the dominant country for both the general websites, FCWs, and PCWs. However, there is a notable difference in the distribution of malicious content, with a higher prevalence in the latter category. This insight highlights the potential need for increased cyber security measures for FCWs and PCWs, especially in the United States. The ``Others" category, while not an individual country, still contributes significantly to the total distribution of websites and malicious content.

\begin{table*}[t]
\caption{An overview of the distribution of the (top-1M, FCWs, and PCWs) across different countries. The names are coded using Alpha-3, where GBR stands for the United Kingdom, which here includes Northern Ireland.}\label{tab:MillionCountries}\label{tab:FCWCountries}\label{tab:PCWCountries}
\begin{minipage}{0.32\linewidth}
\centering
\caption*{General Websites}
\scalebox{0.8}{\begin{tabular}{lrrrrr}
\Xhline{1\arrayrulewidth}
\Xhline{1\arrayrulewidth}
Country & \# & \% & MC & MPCP & MP\\
\Xhline{1\arrayrulewidth}
USA  & 941 & \dd{46}{45.75} & 44 & \cc{5}{4.68} & \cc{2}{2.14} \\
DEU & 146 & \dd{7}{7.10} & 3 & \cc{2}{2.05} & \cc{1}{0.15} \\
FRA & 101 & \dd{5}{4.91} & 6 & \cc{6}{5.94} & \cc{1}{0.29} \\
CHN & 62 & \dd{3}{3.01} & 2 & \cc{3}{3.23} & \cc{1}{0.10} \\
NLD & 62 & \dd{3}{3.01} & 2 & \cc{3}{3.23} & \cc{1}{0.10} \\
CAN & 51 & \dd{3}{2.48} & 2 & \cc{3}{3.92} & \cc{1}{0.10} \\
GBR & 40 & \dd{2}{1.94} & 0 & 0 & 0 \\
AUS & 24 & \dd{1}{1.17} & 3 & \cc{13}{12.50} & \cc{1}{0.15} \\
IRL & 19 & \dd{1}{0.92} & 1 & \cc{5}{5.26} & \cc{1}{0.05} \\
BEL & 4 & \dd{1}{0.19} & 0 & 0 & 0 \\
Otr. & 607 & \dd{30}{29.51} & 121 & \cc{20}{19.93} & \cc{6}{5.88} \\
\Xhline{1\arrayrulewidth}
Total & 2057 & \dd{100}{100} & 92 & \dd{5}{4.47} & \cc{5}{4.47}\\
\Xhline{1\arrayrulewidth}
\Xhline{1\arrayrulewidth}
\end{tabular}}
\end{minipage}~
\begin{minipage}{0.32\linewidth}
\centering
\caption*{FCWs}
\scalebox{0.8}{\begin{tabular}{lrrrrr}
\Xhline{1\arrayrulewidth}
\Xhline{1\arrayrulewidth}
Country & \# & \% & MC & MPCP & MP\\
\Xhline{1\arrayrulewidth}
USA  & 399 & \dd{51}{50.63} & 171 & \cc{43}{42.86} & \cc{22}{21.70} \\
BEL & 88 & \dd{11}{11.17} & 62 & \cc{100}{70.45} & 7.87 \\
DEU & 74 & \dd{9}{9.39} & 16 & \cc{22}{21.62} & \cc{2}{2.03}\\
NLD & 55 & \dd{7}{6.98} & 17 & \cc{31}{30.91} & \cc{2}{2.16} \\
AUS & 42 & \dd{5}{5.33} & 10 & \cc{24}{23.81} & \cc{1}{1.27} \\
FRA & 20 & \dd{3}{2.54} & 13 & \cc{65}{65} & \cc{2}{1.65} \\
GBR & 17 & \dd{2}{2.16} & 9 & \cc{53}{52.94} & \cc{1}1.14 \\
RUS & 13 & \dd{2}{1.65} & 2 & \cc{15}{15.38} & \cc{1}{0.25} \\
CAN & 8 & \dd{1}{1.02} & 0 & 0 & 0 \\
ROU & 7 & \dd{1}{0.89} & 3 & \cc{43}{42.86} & \cc{1}{0.38} \\
Otr. & 65 & \dd{8}{8.25} & 16 & \cc{25}{24.62} & \cc{2}{2.03} \\
\Xhline{1\arrayrulewidth}
Total & 788 & \dd{100}{100} & 319 & \cc{41}{40.48} & \cc{41}{40.48}\\
\Xhline{1\arrayrulewidth}
\Xhline{1\arrayrulewidth}
\end{tabular}}
\end{minipage}~
\begin{minipage}{0.32\linewidth}
\centering
\caption*{PCWs}
\scalebox{0.8}{\begin{tabular}{lrrrrr}
\Xhline{1\arrayrulewidth}
\Xhline{1\arrayrulewidth}
Country & \# & \% & MC & MPCP & MP\\
\Xhline{1\arrayrulewidth}
USA  & 485 & \dd{67}{67.27} & 126 & \cc{26}{25.98} & \cc{18}{17.48} \\
NLD & 40 & \dd{6}{5.55} & 2 & \cc{5}{5} & \cc{1}{0.28} \\
CHN & 28 & \dd{4}{3.88} & 6 & \cc{21}{21.43} & \cc{1}{0.83} \\
GBR & 22 & \dd{3}{3.05} & 3 & \cc{14}{13.64} & \cc{1}{0.42} \\
IRL & 21 & \dd{3}{2.91} & 1 & \cc{4}{4.76} & \cc{1}{0.14} \\
CAN & 16 & \dd{2}{2.22} & 3 & \cc{19}{18.75} & \cc{1}{0.42} \\
IND & 16 & \dd{2}{2.22} & 2 & \cc{13}{12.50} & \cc{1}{0.28} \\
DEU & 15 & \dd{2}{2.08} & 1 & \cc{7}{6.67} & \cc{1}{0.14} \\
FRA & 15 & \dd{2}{2.08} & 2 & \cc{13}{13.33} & \cc{1}{0.28} \\
BEL & 11 & \dd{2}{1.53} & 5 & \cc{46}{45.45} & \cc{1}{0.69} \\
Otr. & 52 & \dd{7}{7.21} & 9 & \cc{17}{17.31} & \cc{1}{1.25} \\
\Xhline{1\arrayrulewidth}
Total & 721 & \dd{100}{100} & 160 & \dd{22}{22.19} & \dd{22}{22.19}\\
\Xhline{1\arrayrulewidth}
\Xhline{1\arrayrulewidth}
\end{tabular}}
\end{minipage}
\end{table*}

\BfPara{Free Websites} 
Table~\ref{tab:FCWCountries} displays the distribution of FCWs across the top hosting countries. The United States leads the list, hosting 50.6\% of the FCWs, followed by Belgium (11.2\%), Germany (9.4\%), the Netherlands (7\%), and Australia (5.3\%). France, the United Kingdom, Russia, Canada, and Romania have smaller percentages of hosted FCWs. In terms of MC, the United States has the highest count (171), while Belgium has the highest MPCP at 70.5\%. The overall MP for FCWs is 40.5\%.

The data reveals that the United States is the dominant hosting country for FCWs, with over half of the websites hosted there. However, Belgium has the highest proportion of malicious content, as indicated by the MPCP value. This information highlights the need for increased security measures for FCWs, particularly in countries with higher concentrations of malicious content. The ``Others" category, which collectively represents 8.3\% of the FCWs, also contributes significantly to the total distribution of malicious content, with an MC of 16 and an MP of 2\%.

\BfPara{Premium Websites} 
Table~\ref{tab:PCWCountries} shows the distribution of PCWs across the top hosting countries. The United States dominates the list, hosting 67.3\% of the PCWs. The Netherlands, China, the United Kingdom, and Ireland follow with smaller percentages of 5.6\%, 3.9\%, 3.1\%, and 2.9\%, respectively. Canada, India, Germany, France, and Belgium also host a minor portion of PCWs. In terms of MC, the United States has the highest count at 126, but Belgium has the highest MPCP at 45.5\%.

The overall MP for PCWs is 22.2\%. The data highlights the significant concentration of PCWs in the United States, but it also shows that Belgium has a higher proportion of malicious content in its PCWs, as evidenced by its MPCP value. This suggests that security measures should be strengthened for PCWs, especially in countries with a higher concentration of malicious content. The ``Others" category, which collectively represents 7.2\% of the PCWs, also contributes significantly to the total distribution of malicious content, with an MC of 9 and an MP of 1.3\%.

\BfPara{Free Websites versus Premium Websites} 
A comparison between the distribution of FCWs and PCWs across the top hosting countries reveals several insights, as shown in Table ~\ref{tab:FCWCountries}. The United States is the top hosting country for both types of websites, with 50.6\% of FCWs and 67.3\% of PCWs. However, the distribution of FCWs is more spread across various countries, with Belgium (11.2\%), Germany (9.4\%), and the Netherlands (7\%) hosting a substantial percentage of FCWs. On the other hand, the distribution of PCWs is more concentrated in the United States, with other countries like the Netherlands, China, and the United Kingdom hosting smaller percentages of 5.6\%, 3.9\%, and 3.1\%, respectively. 

The FCWs have a higher overall MC of 319 compared to PCWs with 160. The MP for FCWs is significantly higher at 40.5\% compared to PCWs, which have an MP of 22.2\%. This suggests that FCWs are more likely to contain malicious content than PCWs. Belgium has the highest MPCP for both FCWs (70.5\%) and PCWs (45.5\%). This suggests that despite hosting a smaller percentage of websites, Belgium has a high proportion of malicious content.

Overall, the distribution of FCWs is more dispersed across various countries than PCWs, primarily concentrated in the United States. The malicious content rates in both types of websites highlight the need for improved security measures, particularly in countries with a high concentration of malicious content.

\subsection{Per-Category Country-level Analysis}
Consistent with the prior work that initiated this line of work, we explore the geographical distribution of the free and premium content websites across the various analysis dimensions by considering their category type, emphasizing the type of content such websites serve. Namely, the contents are divided into books, games, movies, music, and software websites. 

\BfPara{Books Websites} 
Table ~\ref{tab:bookCountries} shows the distribution of free and premium book content websites across different countries. For FCWs, the United States has the highest share, accounting for roughly 58\% of the total, followed by Germany at roughly 8\%, Belgium at roughly 6\%, and Australia at roughly 4\%. The United States also has the highest MC with 32 instances, while Belgium has the highest MPCP at roughly two-thirds of the total. Furthermore, the United States leads in MP with 22.2\%.

For PCWs, the United States maintains dominance with roughly 62\%, followed by China, Canada, the Netherlands, and the United Kingdom. We also note that the United States has the highest MC with 34 instances, while Canada and the United Kingdom have the highest MPCP at roughly 38\%. Once again, the United States has the highest MP, at roughly 18\%. Comparing FCWs and PCWs, the United States dominates both types of websites. Belgium has a significantly high MPCP in FCWs. Moreover, China, Canada, and the United Kingdom have considerable MPCP values in PCWs. 

In summary, there is a significant and clear difference in malicious content distribution in FCWs and PCWs, with slightly higher MP values found in FCWs than PCWs.

\begin{table*}[t]
\caption{An overview of the distribution per category (FCWs vs. PCWs, books, games) across different countries.}\label{tab:FPCountries}\label{tab:bookCountries}\label{tab:gamesCountries}
\begin{minipage}{0.32\linewidth}
\centering
\caption*{Overall}
\scalebox{0.80}{\begin{tabular}{lrrrrr}
\Xhline{1\arrayrulewidth}
\Xhline{1\arrayrulewidth}
Country & \# & \% & MC & MPCP & MP\\
\Xhline{1\arrayrulewidth}
USA & 884 & \dd{59}{58.47} & 297 & \cc{34}{33.60} & \cc{20}{19.64} \\
BEL & 99 & \dd{7}{6.55} & 67 & \cc{68}{67.68} & \cc{4}{4.43} \\
NLD  & 95 & \dd{6}{6.28} & 19 & \cc{20}{20} & \cc{1}{1.26} \\
DEU & 89 & \dd{6}{5.89} & 17 & \cc{19}{19.10} & \cc{1}{1.12} \\
AUS & 48 & \dd{3}{3.17} & 10 & \cc{21}{20.83} & \cc{1}{0.66} \\
GBR & 39 & \dd{3}{2.58} & 12 & \cc{31}{30.77} & \cc{1}{0.08} \\
FRA & 35 & \dd{2}{2.31} & 15 & \cc{43}{42.86} & \cc{1}{0.99} \\
CHN & 33 & \dd{2}{2.18} & 7 & \cc{21}{21.21} & \cc{1}{0.46} \\
CAN & 24 & \dd{2}{1.59} & 3 & \cc{13}{12.50} & \cc{1}{0.20} \\
IRL & 22 & \dd{2}{1.46} & 1 & \cc{5}{4.55} & \cc{1}{0.07} \\
IND & 18 & \dd{2}{1.19} & 3 & \cc{28}{16.67} & \cc{1}{0.20} \\
RUS & 15 & \dd{2}{0.99} & 2 & \cc{24}{13.33} & \cc{1}{0.13} \\
FIN & 12 & \dd{1}{0.8} & 1 & \cc{39}{8.33} & \cc{1}{0.07} \\
SGP & 10 & \dd{1}{0.66} & 1 & \cc{16}{1} & \cc{1}{0.07} \\
Otr. & 86 & \dd{9}{5.7} & 24 & \cc{22}{27.91} & \cc{2}{1.59} \\
\Xhline{1\arrayrulewidth}
Total & 1509 & \dd{100}{100} & 479 & \cc{32}{31.75} & \cc{32}{31.75}\\
\Xhline{1\arrayrulewidth}
\Xhline{1\arrayrulewidth}
\end{tabular}}
\end{minipage}~
\begin{minipage}{0.32\linewidth}
\centering
\caption*{Books}
\scalebox{0.80}{\begin{tabular}{lrrrrr}
\Xhline{1\arrayrulewidth}
\Xhline{1\arrayrulewidth}
\multicolumn{6}{c}{Free Content Websites}\\
\Xhline{1\arrayrulewidth}
Country & \# & \% & MC & MPCP & MP\\
\Xhline{1\arrayrulewidth}
USA  & 84 & \dd{58}{58.33} & 32 & \cc{38}{38.10} & \cc{22}{22.22} \\
DEU & 11 & \dd{8}{7.64} & 0 & 0 & 0 \\
BEL & 9 & \dd{6}{6.25} & 6 & \cc{67}{66.67} & \cc{4}{4.17} \\
AUS & 6 & \dd{4}{4.17} & 0 & 0 & 0 \\
FRA & 4 & \dd{3}{2.78} & 2 & \cc{50}{50} & \cc{1}{1.39} \\
Otr. & 30 & \dd{21}{20.83} & 3 & \cc{10}{10} & \cc{2}{2.08} \\
\Xhline{1\arrayrulewidth}
Total & 144 & \dd{10}{9.54} & 43 & \cc{30}{29.86} & \cc{30}{29.86} \\
\Xhline{1\arrayrulewidth}
\Xhline{1\arrayrulewidth}
\multicolumn{6}{c}{Premium Content Websites}\\
\Xhline{1\arrayrulewidth}
USA  & 118 & \dd{62}{61.78} & 34 & \cc{29}{28.81} & \cc{18}{17.80} \\
CHN & 13 & \dd{7}{6.81} & 4 & \cc{31}{30.77} & \cc{2}{2.09} \\
CAN & 8 & \dd{4}{4.19} & 3 & \cc{38}{37.50} & \cc{2}{1.57} \\
NLD & 8 & \dd{4}{4.19} & 1 & \cc{13}{12.50} & \cc{1}{0.52} \\
GBR & 8 & \dd{4}{4.19} & 3 & \cc{38}{37.50} & \cc{2}{1.57} \\
Otr. & 36 & \dd{19}{18.85} & 8 & \cc{22}{22.22} & \cc{4}{4.19} \\
\Xhline{1\arrayrulewidth}
Total & 191 & \dd{100}{100} & 53 & \cc{28}{27.75} & \cc{28}{27.75}\\
\Xhline{1\arrayrulewidth}
\Xhline{1\arrayrulewidth}
\end{tabular}}
\end{minipage}~
\begin{minipage}{0.32\linewidth}
\centering
\caption*{Games}
\scalebox{0.80}{\begin{tabular}{lrrrrr}
\Xhline{1\arrayrulewidth}
\Xhline{1\arrayrulewidth}
\multicolumn{6}{c}{Free Content Websites}\\
\Xhline{1\arrayrulewidth}
Country & \# & \% & MC & MPCP & MP\\
\Xhline{1\arrayrulewidth}
USA  & 39 & \dd{50}{50} & 31 & \cc{80}{79.49} & \cc{40}{39.74} \\
BEL & 10 & \dd{13}{12.82} & 10 & \cc{100}{100} & \cc{13}{12.82} \\
MDA & 5 & \dd{6}{6.41} & 0 & 0 & 0 \\
NLD & 4 & \dd{5}{5.13} & 3 & \cc{75}{75} & \cc{4}{3.85} \\
ROU & 3 & \dd{4}{3.85} & 0 & 0 & 0 \\
Otr. & 17 & \dd{22}{21.79} & 6 & \cc{35}{35.29} & \cc{8}{7.69} \\
\Xhline{1\arrayrulewidth}
Total & 78 & \dd{100}{100} & 50 & \cc{64}{64.10} & \cc{64}{64.10} \\
\Xhline{1\arrayrulewidth}
\Xhline{1\arrayrulewidth}
\multicolumn{6}{c}{Premium Content Websites}\\
\Xhline{1\arrayrulewidth}
USA  & 67 & \dd{60}{60.36} & 31 & \cc{46}{46.27} & \cc{28}{27.93} \\
NLD & 13 & \dd{12}{11.71} & 0 & 0 & 0 \\
GBR & 5 & \dd{5}{4.50} & 0 & 0 & 0 \\
CHN & 4 & \dd{4}{3.60} & 0 & 0 & 0 \\
FRA & 4 & \dd{4}{3.60} & 0 & 0 & 0 \\
Otr. & 18 & \dd{16}{16.22} & 4 & \cc{22}{22.22} & \cc{4}{3.60} \\
\Xhline{1\arrayrulewidth}
Total & 111 & \dd{100}{100} & 35 & \cc{32}{31.53} & \cc{32}{31.53}\\
\Xhline{1\arrayrulewidth}
\Xhline{1\arrayrulewidth}
\end{tabular}}
\end{minipage}
\end{table*}

\BfPara{Games Websites} 
Table ~\ref{tab:gamesCountries} summarizes the distribution for the games category across countries. For FCWs, the United States has almost half, followed by Belgium at almost 13\%, then Moldova at almost 6\%, and the Netherlands at almost 5\%. The United States has the highest MC with 31 instances and the highest MPCP at almost 80\%. Notably, Belgium stands out with the highest MPCP of 100\% and a MP of almost 13\%. Moreover, the United States leads in MP with almost 40\%. 

As for PCWs, the United States dominates once again with more than 60\%, followed by the Netherlands, the United Kingdom, China, and France. In terms of MP, the United States leads with almost 28\%, while other countries such as the United Kingdom, China, and France---surprisingly---have no reported malicious instances.

In summary, the United States leads in both free and premium game contents websites, with a higher percentage of malicious content in FCWs. Belgium also has a significant presence in FCWs, with a striking 100\% MPCP rate. Other countries have a lesser contribution, and others have no malicious content reported in either FCWs or PCWs.

\BfPara{Movies Websites} 
Table ~\ref{tab:movieCountries} shows the distribution for the movies websites across the studied dimension in different countries. In the case of FCWs, the United States has the largest share with about 47\% of websites, followed by Germany at roughly 15\%, Australia at roughly 10\%, and the Netherlands at roughly 9\%. The United States has the highest MC of 34 and an MPCP of about 23\%

For MPCP, we found that Belgium has the highest MPCP, at 31.8\%. Moreover, the United States is shown to have the highest MP at 11\%. Similarly, for PCWs, we found that the United States dominates with more than 77\%, followed by the Netherlands, China, and Ireland, each of which has only around 4\%--5\%. The United States again has the highest MC with 20 instances and an MPCP of 17\%. China and the Netherlands exhibit similar MPCP values distribution with around 14\% and 13\%, respectively. Moreover, the United States leads in MP at around only 13\%, while most other countries have a smaller number of malicious instances. When comparing FCWs and PCWs, it is evident that the United States has a more significant share of both types of websites.

In summary, the United States has a higher MP in PCWs compared to FCWs, while the MC and MPCP are lower in PCWs. Moreover, the highest MPCP value for a country in FCWs is observed in Belgium, compared to more evenly distributed values in PCWs among countries like China and the Netherlands. This suggests that there may be a difference in the distribution of malicious content between FCWs and PCWs for this category.

\begin{table*}[t]
\caption{An overview of the distribution per category (movies, music, and software) across different countries.}\label{tab:movieCountries}\label{tab:musicCountries}\label{tab:softwareCountries}
\begin{minipage}{0.32\linewidth}
\centering
\caption*{Movies}
\scalebox{0.8}{\begin{tabular}{lrrrrr}
\Xhline{1\arrayrulewidth}
\Xhline{1\arrayrulewidth}
\multicolumn{6}{c}{Free Content Websites}\\
\Xhline{1\arrayrulewidth}
Country & \# & \% & MC & MPCP & MP\\
\Xhline{1\arrayrulewidth}
USA  & 146 & \dd{47}{47.10} & 34 & \cc{23}{23.29} & \cc{11}{10.97} \\
DEU & 46 & \dd{15}{14.84} & 12 & \cc{26}{26.09} & \cc{4}{3.87} \\
AUS & 30 & \dd{10}{9.68} & 8 & \cc{27}{26.67} & \cc{3}{2.58} \\
NLD & 29 & \dd{9}{9.35} & 8 & \cc{28}{27.59} & \cc{3}{2.58} \\
BEL & 22 & \dd{7}{7.10} & 7 & \cc{32}{31.82} & \cc{2}{2.26} \\
Otr. & 37 & \dd{12}{11.94} & 13 & \cc{35}{35.14} & \cc{4}{4.19} \\
\Xhline{1\arrayrulewidth}
Total & 310 & \dd{100}{100} & 82 & \cc{27}{26.45} & \cc{27}{26.45} \\
\Xhline{1\arrayrulewidth}
\Xhline{1\arrayrulewidth}
\multicolumn{6}{c}{Premium Content Websites}\\
\Xhline{1\arrayrulewidth}
USA  & 118 & \dd{78}{77.63} & 20 & \cc{17}{16.95} & \cc{13}{13.16} \\
NLD & 8 & \dd{5}{5.26} & 1 & \cc{13}{12.50} & \cc{1}{0.66} \\
CHN & 7 & \dd{5}{4.61} & 1 & \cc{14}{14.29} & \cc{1}{0.66} \\
IRL & 6 & \dd{4}{3.95} & 0 & 0 & 0 \\
AUS & 2 & \dd{1}{1.32} & 0 & 0 & 0 \\
Otr. & 11 & \dd{7}{7.24} & 1 & \cc{9}{9.09} & \cc{1}{0.66} \\
\Xhline{1\arrayrulewidth}
Total & 152 & \dd{100}{100} & 23 & \cc{15}{15.13} & \cc{15}{15.13}\\
\Xhline{1\arrayrulewidth}
\Xhline{1\arrayrulewidth}
\end{tabular}}
\end{minipage}~
\begin{minipage}{0.32\linewidth}
\centering
\caption*{Music}
\scalebox{0.8}{\begin{tabular}{lrrrrr}
\Xhline{1\arrayrulewidth}
\Xhline{1\arrayrulewidth}
\multicolumn{6}{c}{Free Content Websites}\\
\Xhline{1\arrayrulewidth}
Country & \# & \% & MC & MPCP & MP\\
\Xhline{1\arrayrulewidth}
USA  & 43 & \dd{54}{53.75} & 19 & \cc{44}{44.19} & \cc{24}{23.75} \\
DEU & 9 & \dd{11}{11.25} & 3 & \cc{33}{33.33} & \cc{4}{3.75} \\
BEL & 5 & \dd{6}{6.25} & 4 & \cc{80}{80} & \cc{5}{5} \\
NLD & 5 & \dd{6}{6.25} & 1 & \cc{20}{20} & \cc{1}{1.25} \\
CAN & 3 & \dd{4}{3.75} & 0 & 0 & 0 \\
Otr. & 15 & \dd{19}{18.75} & 4 & \cc{27}{26.67} & \cc{5}{5} \\
\Xhline{1\arrayrulewidth}
Total & 80 & \dd{100}{100} & 31 & \cc{39}{38.75} & \cc{39}{38.75} \\
\Xhline{1\arrayrulewidth}
\Xhline{1\arrayrulewidth}
\multicolumn{6}{c}{Premium Content Websites}\\
\Xhline{1\arrayrulewidth}
USA  & 58 & \dd{67}{67.44} & 13 & \cc{22}{22.41} & \cc{15}{15.12} \\
FIN & 4 & \dd{5}{4.65} & 0 & 0 & 0 \\
IRL & 4 & \dd{5}{4.65} & 0 & 0 & 0 \\
NLD & 4 & \dd{5}{4.65} & 0 & 0 & 0 \\
GBR & 3 & \dd{4}{3.49} & 0 & 0 & 0 \\
Otr. & 13 & \dd{15}{15.12} & 2 & \cc{15}{15.38} & \cc{2}{2.33} \\
\Xhline{1\arrayrulewidth}
Total & 86 & \dd{100}{100} & 15 & \cc{17}{17.44} & \cc{17}{17.44}\\
\Xhline{1\arrayrulewidth}
\Xhline{1\arrayrulewidth}
\end{tabular}}
\end{minipage}~
\begin{minipage}{0.32\linewidth}
\centering
\caption*{Software}
\scalebox{0.8}{\begin{tabular}{lrrrrr}
\Xhline{1\arrayrulewidth}
\Xhline{1\arrayrulewidth}
\multicolumn{6}{c}{Free Content Websites}\\
\Xhline{1\arrayrulewidth}
Country & \# & \% & MC & MPCP & MP\\
\Xhline{1\arrayrulewidth}
USA  & 87 & \dd{49}{49.43} & 55 & \cc{63}{63.22} & \cc{31}{31.25} \\
BEL & 42 & \dd{24}{23.86} & 35 & \cc{83}{83.33} & \cc{20}{19.89} \\
NLD & 13 & \dd{7}{7.39} & 5 & \cc{39}{38.46} & \cc{3}{2.84} \\
GBR & 7 & \dd{4}{3.98} & 6 & \cc{86}{85.71} & \cc{3}{3.41} \\
DEU & 6 & \dd{3}{3.41} & 1 & \cc{17}{16.67} & \cc{1}{0.57} \\
Otr. & 21 & \dd{12}{11.93} & 11 & \cc{100}{52.38} & \cc{6}{6.25} \\
\Xhline{1\arrayrulewidth}
Total & 176 & \dd{100}{100} & 113 & \cc{64}{64.20} & \cc{64}{64.20} \\
\Xhline{1\arrayrulewidth}
\Xhline{1\arrayrulewidth}
\multicolumn{6}{c}{Premium Content Websites}\\
\Xhline{1\arrayrulewidth}
USA  & 124 & \dd{69}{68.51} & 28 & \cc{23}{22.58} & \cc{16}{15.47} \\
DEU & 9 & \dd{5}{4.97} & 1 & \cc{11}{11.11} & \cc{1}{0.55} \\
BEL & 7 & \dd{4}{3.87} & 3 & \cc{43}{42.86} & \cc{2}{1.66} \\
NLD & 7 & \dd{4}{3.87} & 0 & 0 & 0 \\
FRA & 6 & \dd{3}{3.31} & 1 & \cc{17}{16.67} & \cc{1}{0.55} \\
Otr. & 28 & \dd{16}{15.47} & 1 & \cc{4}{3.57} & \cc{1}{0.55} \\
\Xhline{1\arrayrulewidth}
Total & 181 & \dd{100}{100} & 34 & \cc{19}{18.78} & \cc{19}{18.78}\\
\Xhline{1\arrayrulewidth}
\Xhline{1\arrayrulewidth}
\end{tabular}}
\end{minipage}
\end{table*}

\BfPara{Music Websites} 
Table~\ref{tab:musicCountries} shows the results of the music websites category distribution across the different countries. From a distribution standpoint, the United States still leads in FCWs and PCWs, accounting for more than 53\% and 67\% of each category, respectively. Germany, Belgium, and the Netherlands also have a considerable presence in the FCWs. Analyzing malicious content, FCWs exhibit a higher MP in countries such as the United States ($\approx$24\%) and Belgium (5\%), with the ``Others" category showing a collective MP of 5\%. In contrast, PCWs have lower malicious content rates in most countries, with the United States having an MP of $\approx$15\% and the collective ``Others" category at 2.3\%. Finland, Ireland, the Netherlands, and the United Kingdom have no reported malicious content in their PCWs.

In summary, the United States is the primary contributor to both free and premium music content websites, with higher MP observed in the FCWs compared to the PCWs. Other countries such as Germany, Belgium, and the Netherlands also contribute significantly to music content distribution, displaying varying patterns of malicious content between FCWs and PCWs.

\BfPara{Software Websites}  
Table~\ref{tab:softwareCountries} shows the distribution of the FCWs and PCWs across countries, again highlighting a lead of the United States at almost 50\% and 69\% in the FCWs and PCWs, respectively. Moreover, Belgium and the Netherlands also have a notable presence in both categories. On the other hand, and for malicious content, the FCWs have a higher MP overall, particularly in the United States ($\approx$31.3\%) and Belgium ($\approx$20\%). In contrast, the PCWs have lower malicious content rates across all countries. For instance, the United States has an MP of $\approx$16\%, followed by Belgium ($\approx$2\%) in PCWs. 

In summary, the United States is a major contributor to both FCWs and PCWs, with a higher percentage of malicious content observed in FCWs. Other countries, such as Belgium and the Netherlands, also contribute significantly with software content, with varying levels of maliciousness across FCWs and PCWs.

\subsection{National Cyber Security Index}
NCSI measures the country-level cyber security maturity, and we use this dimension of analysis to understand if there is any trend in the availability of free content malicious websites in a given country and their association with such an index.

Table~\ref{tab:NCSI} lists the relationship between the MPCP of the leading countries in hosting FCWs and PCWs and their scores on different NCSI criteria. The results reveal that the countries hosting websites marked as malicious, as indicated with a high MPCP and MP, have a varying range of NCSI scores, indicating the limitations in some aspects of the scoring criteria to capture this essential feature (security) of those websites at the country level. For instance, the United States, which has a high MPCP, scored only 20 in the cyber threat analysis and information sharing (CTAI) and protection of digital services (PDS) criteria. On the other hand, a country like Belgium, with 4.4\% of MP and 67.68 of MPCP, had a DDL of 75.3\% and a CTAI of 80\%. With 20\% of the hosted websites in it being malicious, the Netherlands had 83.5\% in DDL, and 57\% in CSPD. Noteworthy, Germany, with 19.1\% MPCP and 1.1\% of the total MP, scored 90.9\% in NCSI, which is a relatively higher rate than the other countries. However, we observe that the same country also had a DDL score of 81.4\%.

We note Australia, the United Kingdom, and Canada, with 20.8\%, 30.8\%, and 12.5\% MPCP, contributed only 0.9\% of the total MP, but scored 20\% in PDS,  78.7\% DDL in Australia, 81.6\% in the United Kingdom and 77.1\% in Canada. However, the CSPD score of Canada and the United Kingdom is 71\%. The same trend applies to other countries, where the results show that an average of 75.33\% of the countries that host FCWs and PCWs only averaged 31.8\% of MPCP and 2.9\% of the MP while scoring an average of 62\% in PDS, and 72\% in CTAI and CSPD, and 77.5\% in DDL.

This insight supports the previous hypothesis that some countries may need to improve their cyber security measures to combat cyber threats effectively as the scores may not be indicative of the level of security in certain categories--such as free content websites security. Moreover, we observed that the highest malicious FCWs and PCWs hosting is in countries performing 20\% in CSPD, CTAI, and PDS, highlighting the importance of these criteria in measuring the country-level security matureness regarding the studied threat.

\begin{table}[t]
\centering
\caption{The distribution of FCWs and PCWs across different countries associated with NCSI scores. Studied distribution characteristics for each country: the count, MPCP, MP, and the NCSI ranking scores.}\label{tab:NCSI} 
\scalebox{0.9}{\begin{tabular}{lrrrrrrrr}
\Xhline{1\arrayrulewidth}
\Xhline{1\arrayrulewidth}
CN & \# & MPCP & MP & NCSI & DDL & CSPD & CTAI & PDS\\
\Xhline{1\arrayrulewidth}

USA & \dd{59}{884} & \cc{34}{33.60} & \cc{20}{19.64} & \oo{35}{64.94} & \oo{18}{82} & 100 & \oo{80}{20} & \oo{80}{20} \\

BEL & \dd{7}{99} & \cc{68}{67.68} & \cc{4}{4.43} & \oo{6}{93.51} & \oo{25}{75.34} & 100 & \oo{20}{80} & 100 \\

NLD & \dd{6}{95} & \cc{20}{20} & \cc{1}{1.26} & \oo{17}{83.12} & \oo{17}{83.48} & \oo{43}{57} & 100 & \oo{20}{80} \\

DEU & \dd{6}{89} & \cc{19}{19.10} & \cc{1}{1.12} & \oo{9}{90.91} & \oo{19}{81.43} & 100 & 100 & 100 \\

AUS & \dd{3}{48} & \cc{21}{20.83} & \cc{1}{0.66} & \oo{34}{66.23} & \oo{21}{78.68} & 100 & 100 & \oo{80}{20} \\

GBR & \dd{3}{39} & \cc{31}{30.77} & \cc{1}{0.08} & \oo{22}{77.92} & \oo{19}{81.55} & \oo{29}{71} & 100 & \oo{80}{20} \\

FRA & \dd{2}{35} & \cc{43}{42.86} & \cc{1}{0.99} & \oo{16}{84.42} & \oo{21}{78.59} & \oo{14}{86} & \oo{20}{80} & \oo{20}{80} \\

CHN & \dd{2}{33} & \cc{21}{21.21} & \cc{1}{0.46} & \oo{48}{51.95} & \oo{39}{60.81} & \oo{86}{14} & \oo{80}{20} & \oo{20}{80} \\

CAN & \dd{2}{24} & \cc{13}{12.50} & \cc{1}{0.20} & \oo{28}{70.13} & \oo{13}{77.09} & \oo{29}{71} & 100 & \oo{80}{20} \\

IRL & \dd{2}{22} & \cc{5}{4.55} & \cc{1}{0.07} & \oo{28}{70.13} & \oo{14}{76.23} & \oo{29}{71} & \oo{80}{20} & 100 \\
\Xhline{1\arrayrulewidth}
AVG & \dd{9}{137} & \cc{32}{31.75} & \cc{1}{2.89}  & \oo{25}{75.33} & \oo{23}{77.52} & \oo{23}{77} & \oo{28}{72} & \oo{38}{62}\\
\Xhline{2\arrayrulewidth}
\end{tabular}}
\end{table}

\section{Discussion}\label{sec:discussion}
\BfPara{Overall Takeaway} The results of the country-level analysis convey answers to {\bf RQ1} and {\bf RQ2}. We found that the majority of the investigated websites are located in the United States, with 33.6\% of FCWs and PCWs, compared to 45.8\% of the general websites. At the same time, a significant number of the studied websites were identified as malicious. Overall, the FCWs, PCWs, and general websites had a heavy-tailed distribution over the top hosting countries. 

Surprisingly, the vast majority of the malicious websites in all three types of websites are mostly hosted in the United States, where the MPCP shows a very high percentage in the case of FCWs for most of the studied countries. In contrast, the highest MPCP in PCWs mostly concentrated around the top hosting countries. Interestingly, the case is different on the general websites, where the highest MPCP appears in the eighth of the top hosting countries, indicating the severity of the FCWs in comparison to the other types of websites where the MP in FCWs 40.5\%, 22.2\% in PCWs, and only 4.5\% in the general websites. 

\BfPara{Per-category Analysis Takeaway} The summary of the category websites analysis holds answers to {\bf RQ3} by showing the distribution of FCWs and PCWs over countries. Again, we found that the United States dominates both FCWs and PCWs across the top hosting countries and various content categories. However, the distribution of FCWs is more spread across various countries, such as Belgium, Germany, and the Netherlands. For MC, MP, and MPCP, the FCWs generally have higher malicious content rates than PCWs, with Belgium exhibiting the highest MPCP for both types of websites. 

We conclude that the United States has the highest MC, while Belgium has a relatively high proportion of malicious content despite hosting a smaller percentage of websites. The results highlight the need for improved security measures, particularly in countries with high concentrations of malicious content in FCWs and PCWs. While such measures are desired, at a minimum, this study directly points out the insufficiency of the accepted standard for characterizing the security matureness of a country in light of a specific domain and calls for revising such a standard for hosting capabilities and associated security. On the positive side, Germany, which is considered the second dominant hosting country in most FCWs except in the game and software categories, had one of the lowest malicious content website hosting scores, captured well in the associated measures. 

\BfPara{NCSI Analysis Takeaway} The derived results indicate an answer to {\bf RQ4}, where we found that the most malicious websites are concentrated in countries that have gotten a lower score, at least in one of the following aspects: DDL, CSPD, CTAI, and PDS. This finding supports the examined hypothesis that the weakness in these cyber security aspects could contribute to the high concentration of malicious content on these countries' websites. Therefore, we recommend prioritizing the development of these aspects to improve the overall cyber security measures of these countries, reduce the number of malicious websites, and increase the network's security.

The distribution of FCWs and PCWs over countries identified the most contributed hosting countries; however, due to the high overlap between malicious and benign websites within these countries, it is essential to investigate other factors which could cause this weakness. It was also found that the most malicious contributed countries have gotten a lower score in at least one aspect, such as Cyber Security Policy Development (CSPD), Cyber Threat Analysis and Information (CTAI), or Protection of Digital Services (PDS). 

\BfPara{Contrast with the Literature} The results of the NCSI analysis show compatibility with Alabduljabbar \etal{'s}~\cite{AlabduljabbarM22} work where we found indicators that the score of the policy development is relatively low in the countries with a high MPCP, in general. Their findings were drawn from examining the privacy policies, where our results are drawn by tracking the security development indicators of the country and the security state of such websites hosted in a country. As such, our study provides other means of supporting the findings of this prior work. 

We also found that the prior work investigated the security of websites and their geographical distribution at the country level and showed variations between the vulnerable websites per country~\cite{BangeraG17, Sandoval-Guzman17, shafqatM16, VaughanZ07, VelasquezE18, VerkijikaW18, WakelingKJKS22, ZareR14, ZhaoZZ10, MacakogluM22}. However, most of these studies focus on e-government, universities, and libraries websites. As such, the distribution of the malicious FCWs has not been discussed in prior work in contrast with the studied in-depth dimensions, although our findings are consistent with some of such literature. Both our findings and the prior work conclude the need for more regulations on the hosting of websites by considering security as an essential criterion, while our work additionally substantiates this need with an evidence-based study that highlights the performance of the existing measures and the gaps that call for further improvements.

\BfPara{Limitations and Recommendations} One of the unexpected results is that malicious FCWs are highly distributed over the hosting countries. This indicates a need to improve to cyber security policies and agreements across those countries to protect users. Moreover, while some of those regulations might be in existence---as indicated by the discrepancy between the NCSI and our measurements, the higher MPCP discovered rate may be due to a lack of {\em enforcement} of such regulations and policies, calling for tracking the enforcement as an equally important aspect of the matureness of the cyber security policy at the country level.  Given the broad usage of the websites class we studied in this paper, it is important to note that our results and findings highlight how cybercrime may transcend borders and nations, making it difficult to contain without a coordinated international collaboration and dialogue, which should be embodied in the nation-level security matureness scoring.

One potential explanation for the United States is the lead in some of the measurements we conducted is that the collection of the websites (FCWs and PCWs) took place from hosts located in the United States, which biases the returned results to only those relevant in the United States---e.g., Google considers a combination of factors to determine the results, including user's location, language, search history, and the relevance of website. We note that such bias would be at the level of the contents, and not unclear whether the infrastructure---the main studied aspect in this study---is taken into account when returning search results. In the future, we will answer this question with further exploration.

While we did not consider the root cause for the maliciousness of those websites---as that is an important yet orthogonal pursuit, it would be interesting to explore that in the future. One potential factor contributing to our results concerning the distribution of malicious websites across countries is perhaps the difference in access restrictions, data privacy laws, or other digital security measures across different regions or nations, which could lead to varying levels of risk when accessing content from those areas (and, by the same token, security assurance).

As a primary recommendation of this paper, and based on the key findings, there is a need to develop a better cyber security policies and regulations to reduce the risk exposure for users who access these websites. Moreover, while this work provides an overview of the country-level distribution patterns associated with FCWs, PCWs, and their association, much work remains to be done to identify the correlation between the maliciousness of a website and its regional environments.

\section{Conclusion and Future Work} \label{sec:final}
We examined the distribution of free and premium content websites across different countries, showing that malicious FCWs are heavily concentrated in some countries and highlighting the need for more mature cyber security policies to ensure security. We also examined the discrepancy between the NCSI scores of hosting countries and malicious FCW averages. The findings presented here can help inform strategies to better protect users against vulnerabilities beyond their control. Our study is not without limitations, including the need to revisit the data collection, annotation, and website types, which can all be continuously improved to provide better coverage, representation, and data balance. In the future, it would also be important to understand through measurements the different types of maliciousness different websites have, and their severity, which may impact the weight of the different policy scores and their relevance.  Identifying the weaknesses exploited in different categories or regions may help shed light on more precise policy recommendations.

\balance
% Generated by IEEEtranS.bst, version: 1.12 (2007/01/11)

%\bibliographystyle{IEEEtranS}
%\bibliography{sample.bib}

\end{document}